\documentclass[aps,prl,amsmath,amssymb,nofootinbib,superscriptaddress,twocolumn]{revtex4-1}
\usepackage{graphicx}
\usepackage{amsmath,pgfmath,pgffor}
\usepackage[usenames, dvipsnames]{color}
\usepackage{indentfirst}
\usepackage{hyperref}
\usepackage{subfigure}
\usepackage{epstopdf}

\usepackage{hyperref}

\newcommand{\be}{\begin{equation}}
\newcommand{\ee}{\end{equation}}

\usepackage{color}

\hypersetup{%
    pdfborder = {0 0 0},
    colorlinks,
    citecolor=blue,
    filecolor=Darkgreen,
    linkcolor=black,
}

\def\lhc{\ell_{\rm hc}}
\def\lat{\ell_{\rm at}}

 \let\b=\beta \let\g=\gamma 
   
 \let\m=\mu   
\let\s=\sigma  \let\f=\varphi 
   
\let\D=\Delta   
\let\Si=\Sigma   
\let\ee=\varepsilon   \let\io=\infty

\def\FF{{\cal F}}

 \def\xx{{\bf x}} \def\yy{{\bf y}}

\def\redv{\bar v}

\def\xx{\mathbf{x}}
\def\de{\mathrm d}

\def\to{\rightarrow} \def\la{\left\langle} \def\ra{\right\rangle}

\newcommand{\beq}{\begin{equation}} \newcommand{\eeq}{\end{equation}}
\newcommand{\wh}{\widehat}

\begin{document}

\title{Microscopic theory of two-step yielding in attractive colloids}

\author{Ada Altieri}
\affiliation{Laboratoire de Physique Th\'eorique, D\'epartement de Physique de l'ENS, \'Ecole normale sup\'erieure, PSL University, Sorbonne Universit\'es, CNRS, 75005 Paris, France}
\author{Pierfrancesco Urbani} 
\affiliation{Institut de Physique Th\'eorique, Universit\'e Paris Saclay, CNRS, CEA, F-91191, Gif-sur-Yvette}
\author{Francesco Zamponi}
\affiliation{Laboratoire de Physique Th\'eorique, D\'epartement de Physique de l'ENS, \'Ecole normale sup\'erieure, PSL University, Sorbonne Universit\'es, CNRS, 75005 Paris, France}

\begin{abstract}
Attractive colloids display two distinct amorphous solid phases: the 
attractive glass, due to particle bonding, and the repulsive glass, due to the hard core repulsion.
By means of a microscopic mean field approach, we analyze their response to a quasi-static shear strain.
We find that the presence of two distinct interaction length scales may result in a sharp two-step yielding process,
which can be associated with a hysteretic stress response, or with a reversible but non-monotonic stress-strain curve.
We derive a generic phase diagram characterized by two distinct yielding lines, an inverse yielding, and a critical point separating the 
hysteretic and reversible regimes. Our results
should be applicable to a large class of glassy materials characterized by two distinct interaction length scales.
\end{abstract}

\maketitle

\emph{Introduction} - 
In the last few years important progress has been achieved in the study of the response of amorphous solids to a shear strain, which can be
characterized by measuring the associated shear stress~\cite{BL11,RTV11,BDBDM17,NFMB17}.
At small enough strain, the stress response is elastic-like, as for standard solids, whereas, upon increasing the strain, the response becomes more complex and characterized by intermittent mesoscopic drops of the stress (``plasticity''), 
while the system remains macroscopically solid~\cite{FL98,RTV11,BL11,BDBDM17,NFMB17,HKLP11,DPSS16,NFMB17}.
At even larger strains, the solid finally yields and starts to flow, which is signaled by the stress becoming, 
on average, independent of the strain~\cite{BDBDM17}.
More precisely, the yielding transition can happen in several ways, depending on the system's properties
and preparation~\cite{dinkgreve2017,BDBDM17}: 
one observes either a very abrupt drop in the shear stress (``brittle behavior'', characteristic of hard molecular and metallic glasses), 
or a smooth crossover when 
the stress reaches a plateau, sometimes displaying an overshoot 
before the onset of flow (``ductile behavior'', characteristic of soft colloidal glasses and
emulsions)~\cite{Kou12,Amann2013,divoux2011}.

Numerous studies have been carried out in order to better characterize the yielding transition, by means of numerical simulations, phenomenological
models, and microscopic approaches.
One interesting outcome is that the yielding transition can be understood, 
in mean field theory, as a spinodal point of the solid phase in presence of disorder~\cite{RUYZ15,RU16,UZ17, BU18}, which has also been confirmed numerically~\cite{JPRS16,PRS17}. 
Once the effect of structural disorder, which results in a distribution of local stresses~\cite{HL98,LLRW14,PVF16}, 
is taken into account, the spinodal can persist leading to brittle behavior, or it can be destroyed by fluctuations,
leading to a ductile behavior~\cite{OBBRT18,PDW18}. 
Nucleation effects can also be studied within the Random First Order Transition theory~\cite{WW12,WW17}.
Mean field theory is thus a good starting point to develop a microscopic theory of yielding, and in particular it provides qualitatively
correct predictions for the behavior of several relevant observables, including the shear modulus, the dilatancy, the onset of plastic behavior, and
the yielding point itself~\cite{RUYZ15,UZ17,JY17,JUZY18}.

The above phenomenology applies to systems whose interaction potential contains a single relevant length scale (e.g. the sphere diameter).
A more complex behavior is observed for potentials with two length scales: a prototypical example is that of attractive colloids with 
square well potentials, characterized by a hard core repulsion with length scale $\lhc$ and an attractive tail with length scale 
$\lat \ll \lhc$~\cite{DL97,PPBEMPSCFP02,Sc02,ST05}. 
These potentials describe, for example, 
a colloidal suspension in the presence of a non-adsorbing polymer \cite{Po2002}. 
A first-principle analysis of these systems has been performed in 
Mode-Coupling Theory (MCT)~\cite{bergenholtz1999,fabbian1999,dawson2000}
and the results confirmed numerically~\cite{foffi2004} and experimentally \cite{PPBEMPSCFP02}. 
They exhibit two distinct amorphous solid phases: a repulsive glass dominated by $\lhc$, where caging is due to the
repulsion of neighboring particles, and an attractive glass dominated by particle bonding over scale $\lat$. 

For these systems, in proximity of the boundary between the two solid phases, 
a complex response to shear has been experimentally observed, characterized by two distinct
yielding transitions~\cite{pham2006yielding,pham2008yielding,KP11,laurati2009structure,laurati2011}: 
the first has been associated to the breaking of attractive bonds, the second to the breaking of repulsion-induced cages.
A similar behavior has been subsequenly observed in other systems characterized by two 
distinct interaction length scales~\cite{chan2012,fernandez2014,shukla2015}.
However, due to the complexity of these experiments, and to the fact that the yielding transition is never
sharply defined in three-dimensional systems, a systematic investigation of this two-step yielding phenomenon in parameter space (density, 
intensity of the attraction, $\lat/\lhc$,
and strain) has not been performed. 
While MCT overall provides a very good description of the rheology of these systems~\cite{Amann2013,PV14,AF14}, it fails to describe
the two-step yielding process; to the best of our knowledge, no other phenomenological or microscopic theory of this 
process has been proposed.

In this letter, we construct a microscopic mean field theory for the 
square well potential and we fully characterize the response of the attractive and repulsive glasses to a quasi-static shear strain,
computing in particular the stress-strain curves and the yielding point of the two glasses. 
We establish a phase diagram in the space of temperature, density, and shear, that unifies 
the usual single yielding of both glasses with the two-step yielding process. We also provide new predictions:
in particular, the existence of an ``inverse spinodal'' for the repulsive glass phase, which results in a hysteretic stress-strain curve,
and of a critical point where this inverse spinodal merges with the yielding of the attractive glass, giving rise to a smooth but 
non-monotonic stress-strain curve.
Our results can be experimentally tested 
 in colloidal-polymer mixtures by tuning the volume fraction, the polymer concentration and the polymer coil radius. 
Using mean field theory as a guide, numerical simulations could also be carried on straightforwardly.

\emph{Model} -
We consider a system of identical $d$-dimensional particles,
in the thermodynamic limit at number density $\rho$ and temperature $T=1/\beta$,
subject to the following two-body square-well potential with attraction strength $U_0$:
\begin{equation}
v(r)=\begin{cases}
\infty & r < \lhc \ , \\
-U_0 & \lhc < r < \lat \ , \\
0 & r > \lat \ .
\end{cases}
\end{equation}
While we are of course ultimately interested in describing systems in $d=3$, we make use here
of the abstract limit of $d\to\infty$ in which both the thermodynamics and dynamics can be exactly 
solved via mean field theory~\cite{PZ10,nature, MKZ16,CKPUZ17}.
The predictions of the theory for what concerns the response to a quasi-static strain in the case $U_0=0$ (pure hard spheres)~\cite{RUYZ15,UZ17,JY17,JUZY18} 
have been extensively tested in $d=3$~\cite{JY17,JUZY18}, and
are qualitatively correct. 
The relevant adimensional control parameters are the scaled packing fraction $\wh\f = 2^d\varphi/d= \rho \pi^{d/2} \lhc^d/[d \, \Gamma \left( 1+d/2 \right)]$, 
the scaled temperature $\widehat{U}_0=\beta U_0=1/\widehat{T}$, and
the scaled interaction length $\lat/\lhc = 1 + \widehat{\sigma}/d$. The control parameters $\widehat\varphi, \widehat T, \widehat \sigma$ have to be kept finite
when $d\to\infty$ to obtain a non-trivial phase diagram.
We also define for convenience a rescaled potential $\redv(y) =\b v[\lhc(1+y/d)]$, such that $\redv(y<0)=\io$, 
$\redv(1<y<\wh\s) = - \wh U_0$, $\redv(y>\wh\s)=0$.

\begin{figure}[t]
\includegraphics[width=\columnwidth]{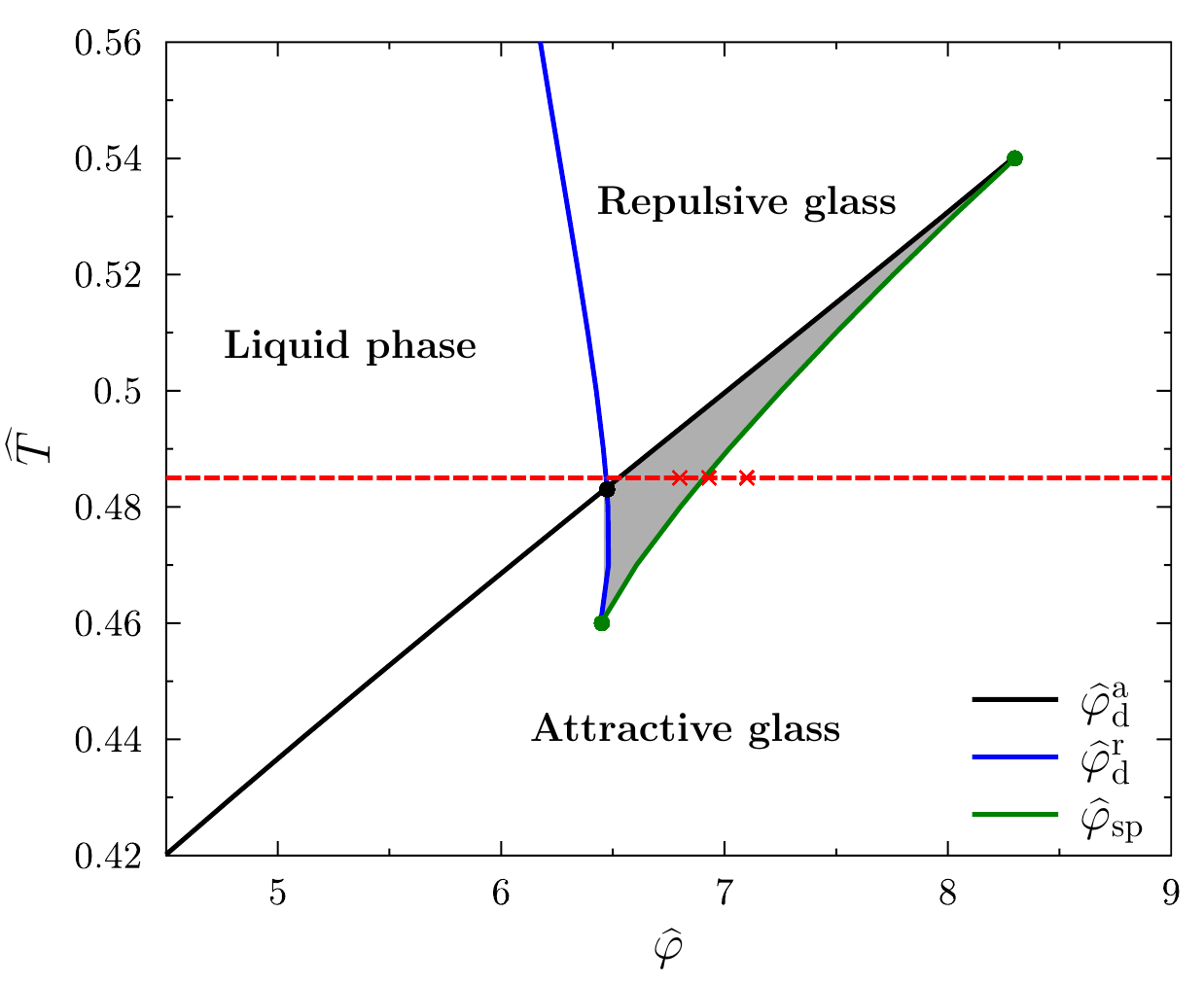}
\caption{Phase diagram in the $(\wh T, \wh \f)$ plane at fixed range $\wh\s=0.062$, in absence of strain. 
The full blue (black) line is the dynamical transition at which the repulsive (attractive) glass appears. 
On the green line, the two glass states merge in a single glass. The glass-glass coexistence region in shaded in gray.
The two green dots indicate higher-order singularities of the Franz-Parisi potential.
The red dashed line indicates the value of $\wh T = 0.485$ studied in the following, and the red crosses indicate the three
state points studied in Fig.~\ref{fig:stress-strain}.
}
\label{phase-diagram}
\end{figure}

\begin{figure*}[t]
\includegraphics[width=0.32\textwidth]{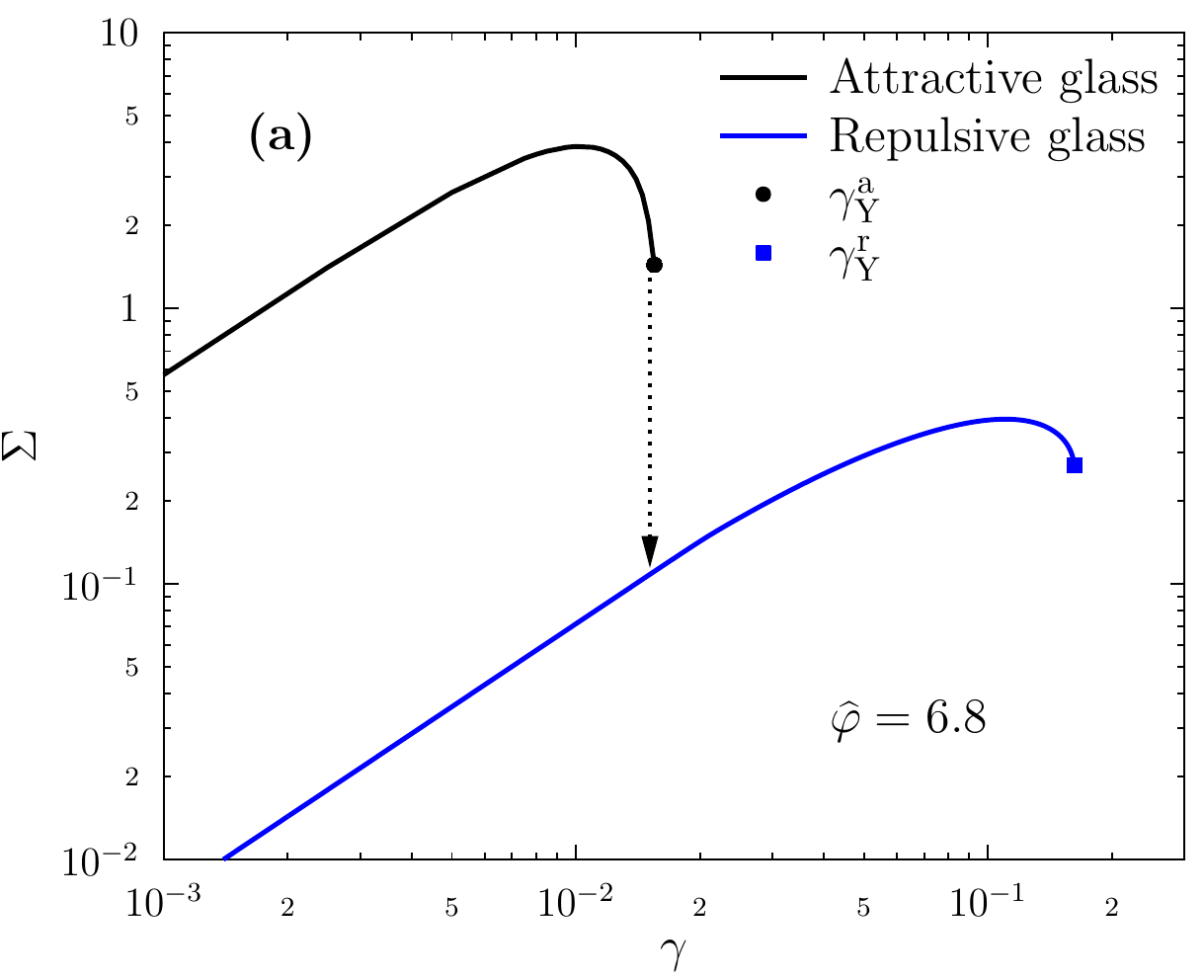}
\includegraphics[width=0.32\textwidth]{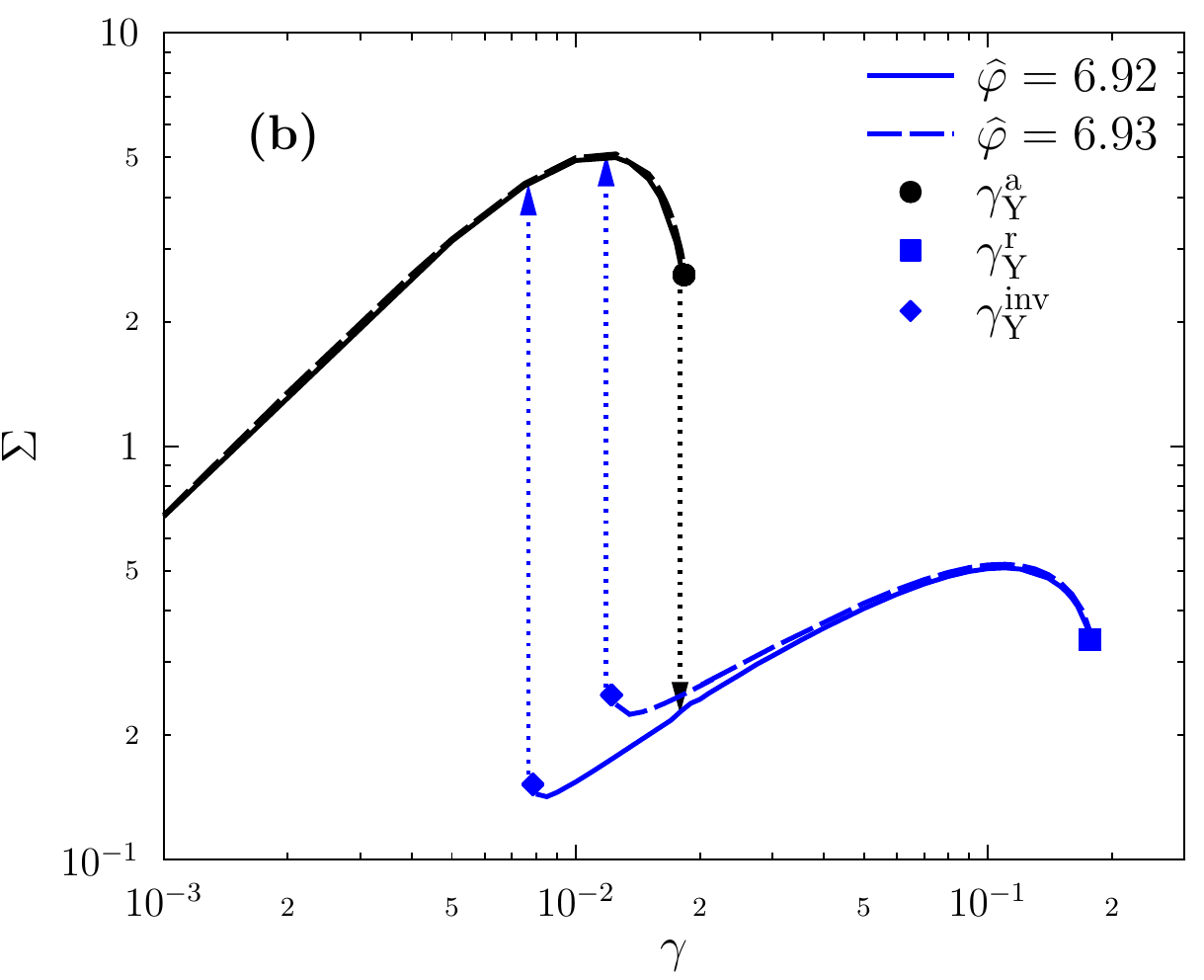}
\includegraphics[width=0.32\textwidth]{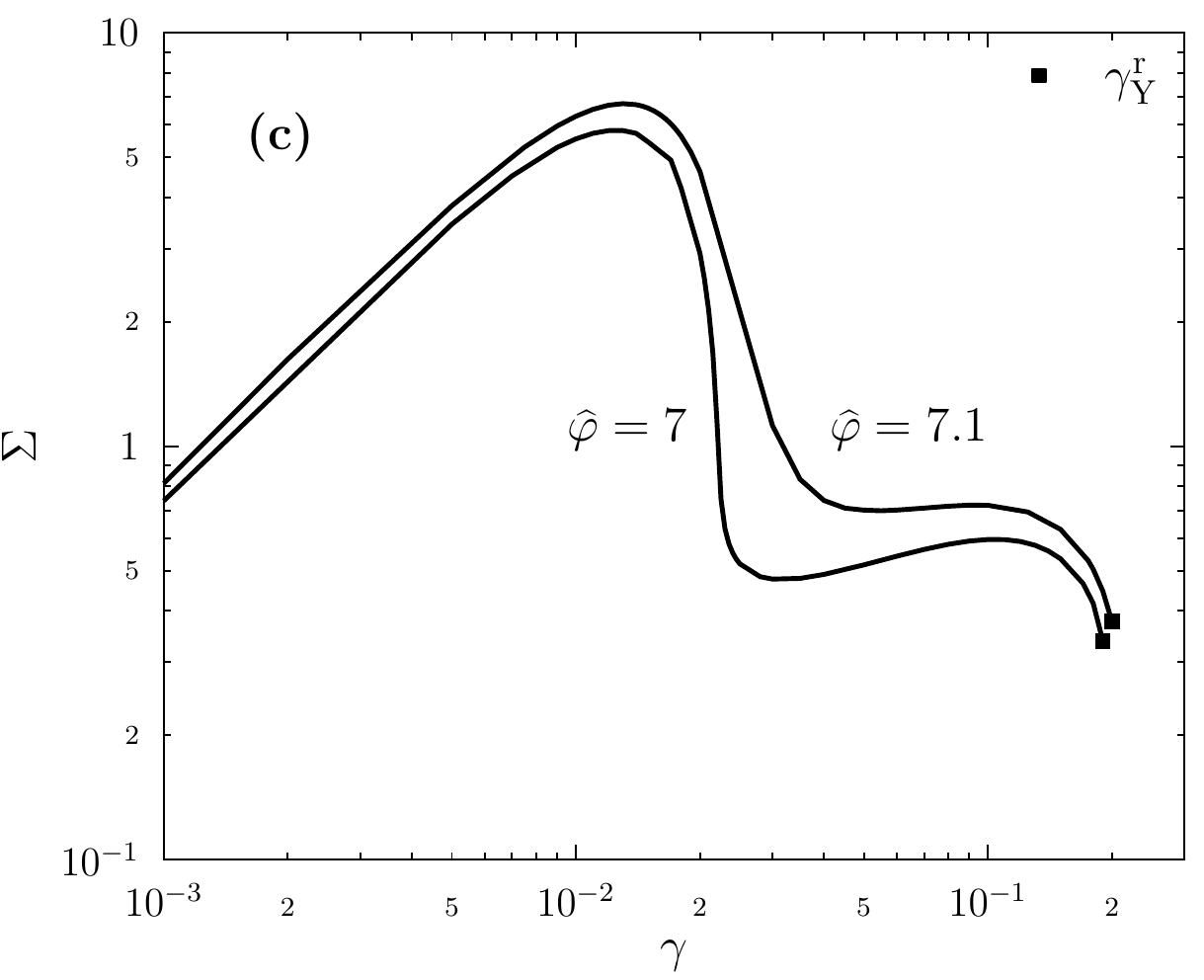}
\caption{
Stress-strain curves in logarithmic scale for several densities $\wh\varphi$, at $\wh T=0.485$ and $\wh\s=0.062$.
(a) Coexistence of the two glasses, displaying distinct yielding points, leading to a two-step yielding process for 
the attractive glass.
(b) The coexistence only happens at $\g>0$, leading to a hysteresis loop at intermediate $\g$, as indicated by the arrows.
(c) The hysteresis loop closes leading to a smooth and reversible, but non-monotonic, stress-strain curve, with a single yielding point.
}
\label{fig:stress-strain}
\end{figure*}

\emph{Glass transition} - 
We first briefly recall the properties of the glass transition in absence of strain~\cite{SZ13}.
The relevant order parameter for the glass transition in $d\to\io$ is the scaled mean square displacement (MSD)
\beq
\Delta(t) = \frac{d}{\lhc^2} \la \frac1N \sum_{i=1}^N | \xx_i(t) - \xx_i(0) |^2 \ra \ .
\eeq
In the liquid phase, $\Delta(t)$ is diffusive at large times, while in the glass phases it reaches a plateau
$\D = \lim_{t\to\io} \D(t)$, which gives the ``cage radius'', i.e. the amplitude of vibrations in the glass phase,
and can be taken as the order parameter for the glass transition.
The value of $\D$ corresponds to the local minimum of the scaled Franz-Parisi potential~\cite{FP95},
given by~\cite{SZ13,RUYZ15,CKPUZ17}:
\begin{equation}\label{FP_potential}
\begin{split}
\mathcal{V}_{\text{FP}}(\D)&= -\log(\D)-\wh{\varphi}\  \mathcal{H}_1 (\D)  \ , \\
\mathcal{H}_1(\D) & = \int_{-\infty}^{\infty} \de y e^{y} \left[ q(\D; y) \log{q(\D;y)} + \redv(y) e^{- \redv(y)} \right]  \ ,  \\
q(\D;y)&= (1-e^{\wh U_0}) \Theta \left( \frac{y+\D-\wh\s}{2 \sqrt{\D} } \right) + e^{\wh U_0} \Theta \left( \frac{y + \D}{2\sqrt{\D} } \right) \ .
\end{split}
\end{equation}
The stationary points of Eq.~(\ref{FP_potential}) are the solutions of:
\begin{equation}
\frac1{\wh\f} = \mathcal{F}_1(\D)= -\D \int_{-\infty}^{\infty} \de y e^y \frac{\partial q(\D;y)}{\partial \D} \log {q(\D;y)} \ .
\label{f1}
\end{equation}
For $\wh U_0=0$ (hard spheres), the function $\FF_1(\D)$ has a single maximum~\cite{PZ10}. 
As a consequence, for 
$\wh\f <  4.8067$, Eq.~\eqref{f1} has
no solutions, corresponding to the liquid phase where the MSD is diffusive. 
For $\wh\f > 4.8067$,  
$\D$ is given by the smallest solution of Eq.~\eqref{f1}, which is a minimum of Eq.~\eqref{FP_potential},
and the dynamics is arrested due to the hard core repulsion (``repulsive glass'').

We now focus on $\wh U_0>0$ (square-well attractive potential) and 
$\wh\s=0.062$, the corresponding phase diagram~\cite{SZ13} being given in Fig.~\ref{phase-diagram}.
For high temperature (low attraction) $\wh T = 1/\wh U_0$, the system is close to the hard sphere limit and a single glass transition to
the repulsive glass phase is observed, which converges to the hard sphere value when $\wh T \to \io$.
At low $\wh T$, a single glass transition is also observed, but at a much smaller value of $\D$, indicating that the glass formation is induced
by the attraction that leads to bonding between particles (``attractive glass'').
At intermediate temperature, the two glass transition lines cross, giving rise to a region where
Eq.~\eqref{f1} has four solutions, two local minima and two local maxima 
of the Franz-Parisi potential, and both glassy solutions formally coexist~\cite{SZ13}, 
the smallest (largest) $\D$ corresponding to the attractive (repulsive) glass.
While the dynamics starting from an equilibrated initial condition should become arrested in the smallest $\D$ solution, i.e.
the attractive glass, other protocols might be able to select the repulsive glass, as we discuss below.
This phenomenology is qualitatively identical to the $d=3$ case~\cite{DL97,Sc02,foffi2004,ST05,PPBEMPSCFP02,bergenholtz1999,fabbian1999,dawson2000}.

\emph{Quasi-static strain} - 
We now briefly recall the ``state following'' construction that allows one to follow the 
evolution of the glass under a quasi-static shear strain~\cite{RUYZ15}.
We assume that the system can be equilibrated at a state point $(\wh\f_g, \wh T_g)$ in the dynamically arrested region of the phase diagram
in Fig.~\ref{phase-diagram}, which corresponds in finite dimensions to the deeply supercooled liquid phase~\cite{RUYZ15,CKPUZ17};
equilibration in this regime
can be nowadays very efficiently achieved by the swap algorithm in numerical simulations~\cite{NBC17}, 
while in experiments one can simply wait long enough
so that the system can overcome the barrier separating distinct glass states and reach equilibrium~\cite{Ca09,BB11}. 
Once an equilibrium configuration~$Y = \{\yy_i\}$ is
obtained at $(\wh\f_g, \wh T_g)$, we assume that a shear strain is applied over a time scale that is slow with respect to all fast relaxation
time scales of the glass (so that it is quasi-static), but fast with respect to the time needed to overcome barriers and relax the glassy structure.

Under these assumptions,
the quasi-static response of the system can be described by the Franz-Parisi construction~\cite{RUYZ15}.
One considers an identical copy of the system,
subject to a weak pinning field to the reference configuration~$Y$, 
which is used as a template for the 
glass state. The resulting Franz-Parisi free energy now depends on two parameters: $\D$, 
the MSD in the glass state, and $\D_r$, the relative MSD of a glass configuration and the reference $Y$~\cite{RUYZ15}. 
Differentiation of the free energy provides two equations for $\D$ and $\D_r$:
\begin{equation}\label{eq:DDr}
\small{
\begin{split}
 & 2 \D_r= \D + \wh\f_g \Delta^2 \int_{-\infty}^{\infty} \de y \; e^y \frac{\partial} {\partial \D} \left[ q_{\gamma} (2 \D_r-\D; y) \log{q(\D; y) } \right] \ , \\
& \frac{2}{\D}= -\wh\f_g \; \int_{-\infty}^{\infty} \de y \; e^y \left( \frac{\partial} {\partial \D_r} q_{\gamma} (2 \D_r-\D; y) \right)\log{q(\D; y) }  \ ,
\end{split}
}
\end{equation}
where
\begin{equation}
q_{\gamma}(\D;y)= \int \frac{d \zeta}{\sqrt{2 \pi}}  e^{-\frac{\zeta^2}2} q (\D + \gamma^2 \zeta^2; y)  \ ,
\end{equation}
and $\g$ is the shear strain. Note that here we work at constant density, but we obtain qualitatively similar results at constant pressure.
These equations can be solved iteratively starting with the initial condition $\D= \D_r = \D(\g=0)$, 
valid at zero strain, and computing numerically the right hand side for a new estimate of $\D$ and $\D_r$, until convergence is reached.
From the converged values of $\D,\D_r$ one can compute the shear stress $\Si$ (scaled in units of temperature $T$, which is natural
in the hard sphere limit), 
\beq\label{eq:Si}
\Si =
\frac{d \wh\f_g}2 \frac{\de}{\de \g} \int_{-\infty}^\infty  \de h\, e^{h} q_\g(2\Delta_r-\Delta; h) \log q(\Delta;  h )  \ ,
\eeq
the pressure, and any other interesting observable~\cite{RUYZ15}.

\emph{Results for the two-step yielding transition} - 
We focus, for illustration, on a fixed potential with $\wh \s = 0.062$ and $\wh T=0.485$ (red line in the phase diagram of Fig.~\ref{phase-diagram}), using
density $\wh\f$ as a control parameter; similar
results are obtained in other regions of the phase diagram.
We examine in particular three values of $\wh\f$ (red crosses in Fig.~\ref{phase-diagram}) for which stress-strain curves are reported in Fig.~\ref{fig:stress-strain}.

For $\g\to 0$, the theory predicts a linear regime $\Si = \m \g$ with shear modulus $\m = 1/\D(\g=0)$~\cite{YZ14}. 
The shear modulus of the attractive glass, which has a much smaller $\D$, is thus much bigger than the one of the repulsive glass. This is
illustrated in Fig.~\ref{fig:stress-strain}a. Upon increasing $\g$, both glasses enter in a non-linear regime, display a stress overshoot, 
and ultimately yield via a spinodal mechanism,
in which the solution of Eq.~\eqref{eq:DDr} is lost via a bifurcation, leading to a square-root singularity of $\Si(\g)$~\cite{RUYZ15}. 
This leads us to define two distinct yielding points $\g_{\rm Y}^{\rm a}, \g_{\rm Y}^{\rm r}$.
In particular, if the system is prepared in the attractive glass state and then strained, it will first undergo a yielding transition  at $\g_{\rm Y}^{\rm a}$ 
towards the repulsive glass, 
which will then yield at $\g_{\rm Y}^{\rm r}$, leading to a two-step yielding process very similar to the one observed in experiments~\cite{pham2006yielding,KP11,laurati2011}.
Note also that if the attractive glass is strained up to $\g_0 \in (\g_{\rm Y}^{\rm a}, \g_{\rm Y}^{\rm r})$, the system yields and jumps to the
repulsive glass. Releasing the strain, the system follows the repulsive glass curve back to $\g=0$, leading to a hysteretic response.
Subsequent straining cycles will follow reversibly the repulsive glass curve. This protocol provides a way to produce the repulsive glass
at this state point, where the dynamics from an equilibrium configuration naturally selects the attractive glass solution.

\begin{figure}[t]
\includegraphics[width=\columnwidth]{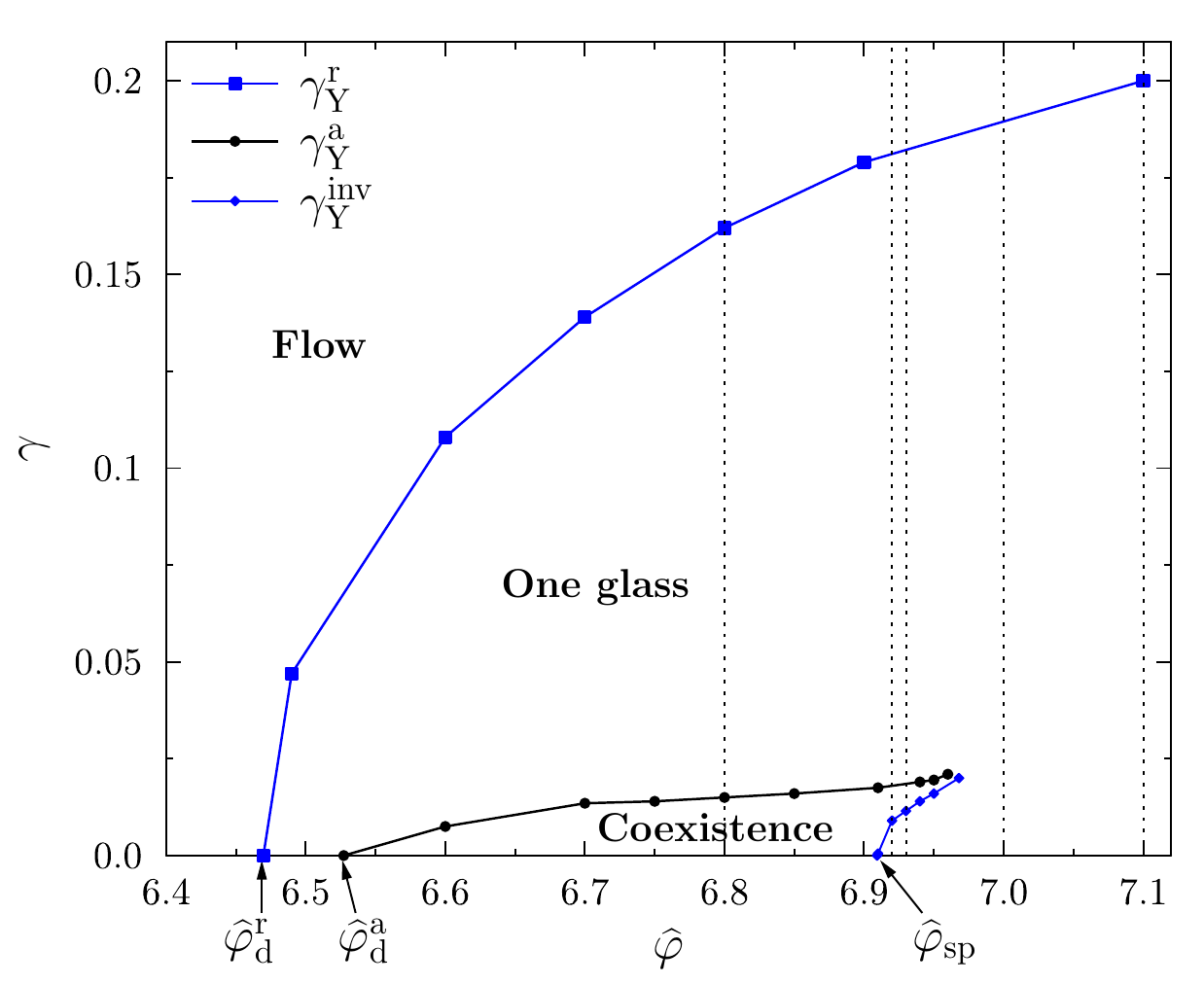}
\caption{Yielding phase diagram in the $(\g,\wh\f)$ plane, obtained by following the evolution in $\wh\f$ of the yielding
points defined in Fig.~\ref{fig:stress-strain}.
The lines $\g_{\rm Y}^{\rm a}$ and $\g_{\rm Y}^{\rm inv}$ delimitate the coexistence region, and merge in a critical point. 
Outside the coexistence region, only one glass exists. The dynamical transitions, at which the $\gamma_{\rm Y}$ vanish,
are respectively located in $\wh \varphi_{\rm d}^{\rm r}=6.4697$, $\wh \varphi_{\rm d}^{\rm a}= 6.5272$, $\wh \varphi_{\rm sp}=6.9093$.
Vertical lines indicated the densities for which the stress-strain curve is reported in Fig.~\ref{fig:stress-strain}.
}
\label{fig:PDfg}
\end{figure}

Upon increasing density (Fig.~\ref{fig:stress-strain}b), 
one reaches the point where at $\g=0$ the two glasses merge into a single one. Beyond that point,
we find --surprisingly-- that the repulsive glass still exists at finite $\g$, and it can be accessed, as before, by increasing
$\g$ beyond the yielding point $\g_{\rm Y}^{\rm a}$.
Upon decreasing $\g$, the repulsive glass 
undergoes an ``inverse yielding'' $\g_{\rm Y}^{\rm inv}$ at which it jumps back into the attractive glass before the strain is fully released. 
In this region we thus find three spinodal points, which
leads to the new prediction of 
a hysteretic response at intermediate $\g$, as illustrated in Fig.~\ref{fig:stress-strain}b.
Note that the hysteresis loop would now be observed during all straining cycles of amplitude $\g_0 \in (\g_{\rm Y}^{\rm a}, \g_{\rm Y}^{\rm r})$.
This prediction could be easily tested in cyclic strain experiments.

Upon further increasing $\wh\f$, the hysteretic loop shrinks and ultimately disappears via a critical point where the repulsive and attractive glasses
merge at finite $\g$. Beyond that point (Fig.~\ref{fig:stress-strain}c), we find a single smooth and reversible stress-strain curve, which however remains
non-monotonic, as a vestige of the hysteretic loop. The stress-strain curve terminates in a single yielding point which is the analytic continuation
of $\g_{\rm Y}^{\rm r}$. 
The evolution with density of the three spinodal points can be reported in a phase diagram in the 
$(\g,\wh\f)$ plane (Fig.~\ref{fig:PDfg}), characterized by a single-glass region, a coexistence region, and a ``flow'' region (where all glasses have become unstable).

\emph{Conclusions} - We studied, within the exact infinite-dimensional solution, an attractive colloid with square-well interaction potential.
In the region of the phase diagram where the attractive and repulsive glasses coexist, 
we detected a two-step yielding mechanism. When the system is prepared in the attractive glass phase, it undergoes a first yielding due to
bond breaking, which brings it to the repulsive glass, followed by a second yielding where the repulsive cage breaks (Fig.~\ref{fig:stress-strain}a). 
Upon increasing density, the repulsive glass becomes unstable at $\g=0$ but remains present at $\g>0$, leading to an unexpected hysteresis
loop at intermediate $\g$, associated to a new ``inverse yielding'' instability of the repulsive glass (Fig.~\ref{fig:stress-strain}b).  
At even higher density, the hysteresis loop closes, leading to a smooth, reversible, but non-monotonic stress-strain curve (Fig.~\ref{fig:stress-strain}c).
We investigated these effects systematically, resulting in a complete phase diagram (Fig.~\ref{fig:PDfg}). 
All the equations needed
to construct the stress-strain curves and the phase diagram are given above, so that the results can be easily extended to other 
ranges of parameters.

We thus proposed a solid theoretical framework to describe the rheology of colloidal systems, 
based on the Franz-Parisi construction and the infinite-dimensional solution.
This work opens new perspectives both for the theoretical, numerical and experimental investigation of these systems.
Theoretically, one could investigate the marginal stability of the glass (the so-called Gardner transition~\cite{nature}), and study how the yielding transitions
are affected by disorder and by a finite shear-rate.
It could be useful to have a more direct comparison between our theoretical predictions in $d\to\io$ 
and numerical simulations in $d=3$, to test the phase diagram of Fig.~\ref{fig:PDfg}. 
Finally, it would be extremely interesting to try to detect experimentally the inverse spinodal we predicted,  
and the associated hysteresis loop. We believe that our results should apply quite generically to systems whose interaction potential displays
two distinct length scales.

\emph{Acknowledgments} - We thank Matthias Fuchs for interesting comments on the manuscript.
This project has received funding from the European Research Council (ERC) under the European Union's Horizon 2020 research and innovation programme (grant agreement 723955 - GlassUniversality).
 This work is supported by ``Investissements d'Avenir'' LabEx PALM (ANR-10-LABX-0039-PALM).
\bibliography{HS}




\end{document}